\documentclass[11pt]{article}
\usepackage{latexsym}  
\usepackage{amssymb}
\usepackage{graphicx}
\usepackage{amsmath}

\begin{document}
\hspace*{11cm} {OU-HET-632/2009}

\begin{center}
{\Large\bf Charged Lepton Mass Spectrum}\\
{\Large\bf and Supersymmetric Yukawaon Model}

\vspace{5mm}
{\bf Yoshio Koide}

{\it Department of Physics, Osaka University,  
Toyonaka, Osaka 560-0043, Japan} \\
{\it E-mail address: koide@het.phys.sci.osaka-u.ac.jp}

\date{\today}
\end{center}

\vspace{3mm}
\begin{abstract}
On the basis of the so-called supersymmetric yukawaon model,
it is investigated what form of the superpotential $W$ can
lead to the observed charged lepton mass spectrum. 
A simple form of $W$ can speculate reasonable values of 
$K(\mu)=(m_e +m_\mu + m_\tau)/(\sqrt{m_e} + \sqrt{m_\mu}
+ \sqrt{m_\tau})^2$ and $\kappa(\mu)=\sqrt{m_e m_\mu m_\tau}/
(\sqrt{m_e} + \sqrt{m_\mu}+ \sqrt{m_\tau})^3$ at an
energy scale $\mu=\Lambda$.
\end{abstract}

\vspace{3mm}

\noindent{\large\bf 1 \ Introduction}

If quarks and leptons are fundamental entities in the nature,
their mass spectra will obey laws of the nature which
have simple forms.
Especially, since the charged lepton masses are almost 
independent of gluon corrections, it will be comparatively easy 
to identify such laws of the nature.
In reference to this problem, for example, we know that an 
empirical formula \cite{Koidemass}
$$
m_e +m_\mu + m_\tau = \frac{2}{3}(\sqrt{m_e} + \sqrt{m_\mu}
+ \sqrt{m_\tau})^2 ,
\eqno(1.1)
$$
is well-satisfied by the observed charged lepton masses
(``pole" mass values).
However, usually, masses which we deal with in a conventional 
mass matrix model are ``running" masses $m_i(\mu)$, not  
pole masses $m_i^{pole}$.
It is also well known that the running masses do not satisfy
the relation (1.1).
In this paper, we will consider that the relation (1.1) is
merely an approximate relation even for the pole masses,
so that we will not attempt to derive the relation (1.1)
as an exact relation.
Instead, we will discuss running masses $m_i(\mu)$ at a 
high energy scale $\mu=\Lambda$.
It is beyond the scope of the present work to explain 
validity of the mass formula (1.1) with the current 
experimental accuracy of order $10^{-5}$; rather we aim 
at accuracy of order $\varepsilon\sim 10^{-3}$.

In the present paper, we investigate the charged lepton mass
spectrum based on the running masses.
Then, for investigating the mass spectrum, it is useful 
to use the following two quantities
$$
K(\mu) \equiv 
\frac{m_e(\mu) +m_\mu (\mu)+m_\tau(\mu)}{(
\sqrt{m_e(\mu)} +\sqrt{m_\mu(\mu)}
+\sqrt{m_\tau(\mu)})^2} ,
\eqno(1.2)
$$
and
$$
\kappa(\mu) \equiv \frac{ \sqrt{m_e(\mu)}
\sqrt{ m_\mu(\mu)} \sqrt{ m_\tau(\mu)}  }{
(\sqrt{m_e(\mu)} +\sqrt{m_\mu(\mu)} 
+\sqrt{m_\tau(\mu)} )^3}  ,
\eqno(1.3)
$$
instead of the mass ratios $m_e(\mu)/m_\tau(\mu)$ 
and $m_\mu(\mu)/m_\tau(\mu)$,
because $K(\mu)$ and $\kappa(\mu)$ are almost independent of 
an energy scale $\mu$ up to the order of $\alpha$ 
\cite{y-mass-0902}
as we show below:
The running masses $m_i(\mu)$ are related to the pole 
masses $m_i^{pole}$ as
$$
 m_i(\mu) = m_i^{pole} \left[ 1 -\frac{\alpha(\mu)}{\pi}
 \left( 1+ \frac{3}{4} \ln \frac{\mu^2}{m_i^2} \right)
 \right] ,
 \eqno(1.4)
$$
under the one-loop approximation \cite{polemass}. 
When we denote $K(\mu)$ and $\kappa(\mu)$ as
$$
K(\mu) = K^{pole} (1+ \varepsilon_K(\mu)) , \ \ \ 
\kappa(\mu) = \kappa^{pole} (1+ \varepsilon_\kappa(\mu)) ,
\eqno(1.5)
$$
where $K^{pole}$ and $\kappa^{pole}$ are defined by 
pole masses correspondingly to $K(\mu)$ and $\kappa(\mu)$, 
respectively, we obtain the deviations $\varepsilon_K(\mu)$ 
and $\varepsilon_\kappa(\mu)$ as follows:
$$
\varepsilon_K(\mu) = \frac{3\alpha(\mu)}{4\pi}  
\sum_i \left( \frac{\sqrt{m_i}}{\sqrt{m_1}+\sqrt{m_2} +\sqrt{m_3}}
 - \frac{ m_i}{m_1 + m_2 +m_3} \right)
\ln \frac{\mu^2}{m_i^2}
 + O(\alpha^2) , 
 \eqno(1.6)
$$
and
$$
\varepsilon_\kappa(\mu)   
= \frac{3\alpha(\mu)}{4\pi}  \sum_i \left( \frac{3}{2} 
\frac{\sqrt{m_i}}{
\sqrt{m_1}+\sqrt{m_2} +\sqrt{m_3}} - \frac{1}{2} 
\right) \ln \frac{\mu^2}{m_i^2} + O(\alpha^2) ,
\eqno(1.7)
$$
respectively, where $m_i=m_i(\mu)$.
Since we can show
$$
\sum_i \left( \frac{\sqrt{m_i}}{\sqrt{m_1}+\sqrt{m_2} +\sqrt{m_3}}
 - \frac{ m_i}{m_1 + m_2 +m_3} \right) =0 ,
 \eqno(1.8)
 $$
 $$
 \sum_i \left( \frac{3}{2} \frac{\sqrt{m_i}}{
\sqrt{m_1}+\sqrt{m_2} +\sqrt{m_3}} 
- \frac{1}{2} \right) =0,
\eqno(1.9)
$$
the values $\varepsilon_K(\mu)$ and $\varepsilon_\kappa(\mu)$ 
are independent of $\mu$ up to the order of $\alpha$.
The numerical results are as follows:
$$
\varepsilon_K(\mu)=  \frac{3\alpha(\mu)}{4\pi} \times 0.98938 
+O(\alpha^2) 
= 1.7236 \times 10^{-3} +O(\alpha^2) ,
\eqno(1.10)
$$
$$
\varepsilon_\kappa(\mu) = \frac{3\alpha(\mu)}{4\pi} \times
(-9.0562)+ O(\alpha^2) = -1.5777 \times 10^{-2}+ O(\alpha^2) .
\eqno(1.11)
$$

\begin{table}
\caption{Energy scale dependence of $K$ and $\kappa$ in a SUSY
model with $\tan\beta=10$: 
$K$ and $\kappa$ for pole masses are given by 
$K^{pole}=0.999989\pm 0.000014$ and
$\kappa^{pole}=(2.0633\pm 0.0001)\times 10^{-3}$.}

\vspace{2mm}

\begin{tabular}{|c|c|c|} \hline
Scale & $\frac{3}{2}K(\mu)$ & $\kappa(\mu)$ \\ \hline
$\mu = M_Z$ & $1.001879 \pm +0.00002$ &
 $(2.02760^{+0.00015}_{-0.00016}) \times 10^{-3}$ \\
$\mu = 10^3$ GeV & $1.00195 \pm 0.00002$ & 
$(2.02710\pm 0.00016) \times 10^{-3}$ \\
$\mu = 10^9$ GeV & $1.00220 \pm 0.00002$ & 
$(2.02469\pm 0.00016) \times 10^{-3}$ \\
$\mu = 10^{12}$ GeV & $1.00230 \pm 0.00002$ & 
$(2.02375^{+0.00017}_{-0.00015}) \times 10^{-3}$ \\
$\mu =2\times 10^{16}$ GeV & $1.00242 \pm 0.00002$ & 
$(2.02268^{+0.00016}_{-0.00015}) \times 10^{-3}$ \\ \hline
\end{tabular}
\end{table}

However, the values (1.10) and (1.11) are pure electromagnetic 
corrections based on Eq.(1.4), and there are corrections from
another diagrams, which are dependent on models and slightly
show $\mu$-dependence.
We show a typical case in a SUSY model with $\tan\beta=10$ in
Table 1, where the input charged lepton mass values are quoted 
from Ref.\cite{qmass-Xing08}.
As seen in Table 1, the quantities $K(\mu)$ and $\kappa(\mu)$
are almost insensitive to the energy scale $\mu$.
From the results of Table 1, we take 
$$
\varepsilon_K(\Lambda) \simeq 2.4 \times 10^{-3} , \ \ \ 
\kappa(\Lambda) \simeq 2.023 \times 10^{-3} ,
\eqno(1.12)
$$
as values which should be explained in this paper.

In reference to this topic, recently, Sumino \cite{Sumino09} 
has proposed an interesting idea that a flavor gauge symmetry 
induces an effect which exactly cancels the deviation of 
$\varepsilon_K(\mu)$ at any energy scale $\mu$.
However, differently from Sumino's idea, in this paper, 
we adopt the standpoint that 
$K^{pole}=2/3$ is accidental.
We consider that our goal is not $K^{pole}=2/3$, but 
$K(\Lambda) =\frac{2}{3}(1+\varepsilon(\Lambda))$ with 
$\varepsilon(\Lambda) \sim 10^{-3}$ from the beginning.
($\varepsilon_K(\mu)$ is deviation of $K(\mu)$ from $K^{pole}$ 
due to running masses, while $\varepsilon(\Lambda)$ is 
deviation of $K(\Lambda)$ from the value $2/3$ in the 
present model.)

In this paper, we try to understand the values (1.12) from 
a new approach to a mass matrix model, the so-called 
``yukawaon model" \cite{yukawaon,e-yukawaon-PRD09}: 
We regard the Yukawa coupling constants $Y_f$ as ``effective"
coupling constants $Y_f^{eff}$ in an effective theory, 
and we consider that 
$Y_f^{eff}$ originate in vacuum expectation values (VEVs)
of new gauge singlet scalars $Y_f$, i.e.
$$
Y_f^{eff} =\frac{y_f}{\Lambda} \langle Y_f\rangle ,
\eqno(1.13)
$$
where $\Lambda$ is a scale of the effective theory which is 
valid at $\mu \leq \Lambda$, and we assume 
$\langle Y_f\rangle \sim \Lambda$.
We refer the fields $Y_f$ as 
``yukawaons" hereafter.  
Note that the effective coupling constants $Y_f^{eff}$ evolve
as in the standard SUSY model below the scale $\Lambda$, since 
a flavor symmetry is completely broken at the high energy scale
$\mu=\Lambda$ (but the supersymmetry is still unbroken at 
$\Lambda > \mu > \Lambda_{SUSY} \sim 10^3$ GeV).

In the yukawaon model, the Higgs scalars are
the same as ones in a conventional model, i.e. we
consider only two Higgs scalars $H_u$ and $H_d$ as an 
origin of the masses (not as an origin of the
mass spectra).
%
For example, we assume an O(3) flavor symmetry 
\cite{Koide-O3-PLB08} and we consider that the yukawaons
$Y_f$ are $({\bf 3}\times{\bf 3})_S= {\bf 1}+{\bf 5}$ 
of O(3)$_F$.
Then, the would-be Yukawa interactions are given by
$$
W_{Y}= \sum_{i,j} \frac{y_u}{\Lambda} u^c_i(Y_u)_{ij} {q}_{j} H_u  
+\sum_{i,j}\frac{y_d}{\Lambda} d^c_i(Y_d)_{ij} {q}_{j} H_d 
$$
$$
+\sum_{i,j} \frac{y_\nu}{\Lambda} \ell_i(Y_\nu)_{ij} \nu^c_{j} H_u  
+\sum_{i,j}\frac{y_e}{\Lambda} \ell_i(Y_e)_{ij} e^c_j H_d +h.c. 
+ \sum_{i,j}y_R \nu^c_i (Y_R)_{ij} \nu^c_j ,
\eqno(1.14)
$$ 
where $q$ and $\ell$ are SU(2)$_L$ doublet fields, and
$f^c$ ($f=u,d,e,\nu$) are SU(2)$_L$ singlet fields.
Here, in order to distinguish each $Y_f$ from others, 
we have assigned U(1)$_X$ charges as 
$Q_X(f^c)=-x_f$, $Q_X(Y_f)= +x_f$ and $Q_X(Y_R)=2x_\nu$.
The VEVs of yukawaons are obtained from supersymmetric
(SUSY) vacuum conditions for a superpotential $W$. 
In the charged lepton sector, we assume
$$
W_e = \lambda_e [\Phi_e \Phi_e \Theta_e] + 
\mu_e [Y_e \Theta_e] 
+W_\Phi , 
\eqno(1.15)
$$
where the fields $\Phi_e$ and $\Theta_e$ have the 
U(1)$_X$ charges $\frac{1}{2} x_e$ and $-x_e$, 
respectively.
Here and hereafter, for convenience, we denote 
a trace ${\rm Tr}[A]$ of a matrix $A$ as $[A]$ 
concisely.
A SUSY vacuum condition $\partial W/\partial \Theta_e=0$
leads to a bilinear mass relation
$$
\langle Y_e\rangle = - \frac{\lambda_e}{\mu_e} 
\langle\Phi_e\rangle \langle\Phi_e\rangle .
\eqno(1.16)
$$
Therefore, the mass spectrum of the charged leptons
are  given by
$$
K = 
\frac{m_e +m_\mu +m_\tau}{(\sqrt{m_e} +\sqrt{m_\mu}
+\sqrt{m_\tau})^2} = \frac{v_1^2+v_2^2+v_3^2}{(
v_1+v_2+v_3)^2} =\frac{[\langle\Phi_e\rangle
\langle\Phi_e\rangle]}{[\langle\Phi_e\rangle]^2} ,
\eqno(1.17)
$$
and
$$
\kappa = \frac{\sqrt{m_e m_\mu m_\tau}}{
(\sqrt{m_e} +\sqrt{m_\mu} +\sqrt{m_\tau} )^3}
=\frac{v_1 v_2 v_3}{(v_1+v_2+v_3)^3} =
 \frac{ \det \langle\Phi_e\rangle}{
[\langle\Phi_e\rangle]^3} ,
\eqno(1.18)
$$
where $\langle\Phi_e\rangle={\rm diag}(v_1,v_2, v_3)$.
We refer the field $\Phi_e$ as an ``ur-yukawaon".
A case with $K=2/3$ (an earlier attempt to give 
$K=2/3$ has been done in Ref.\cite{K-mass90})
is known as the charged lepton mass
formula (1.1) which is excellently satisfied 
by the observed charged lepton masses (pole masses).  
However, note that masses  which we deal with 
in the present paper are running masses, 
so that our goal is not $K^{pole}=2/3$, but 
$K(\Lambda)=(2/3)(1+\varepsilon(\Lambda))$ with 
$\varepsilon(\Lambda) \sim 10^{-3}$.

The term $W_\Phi$ in the superpotential (1.15) has been 
introduced in order to fix a VEV spectrum of 
$\langle\Phi_e\rangle$.
The explicit form is given in the next section.
The purpose of the present paper is heuristically to 
find a form of $W_\Phi$ which can simultaneously 
give reasonable values of $K(\Lambda)$ and $\kappa(\Lambda)$, 
and it is not
to investigate an origin of such the form of $W_\Phi$.
The investigation of the theoretical origin will be our 
next step of the investigations.

\vspace{3mm}

\noindent{\large\bf 2 \ VEV structure of the ur-yukawaon}

In this paper, we consider a model with  explicit U(1)$_X$ 
symmetry breakings of the order $\varepsilon_{SB}$ which is 
negligibly small. 
A prototype of such a model has been discussed in 
Ref.\cite{y-mass-0902}.
We know that, in order to give $K=2/3$ , 
it is essential to bring interaction terms with a traceless 
part $\hat{\Phi}_e \equiv \Phi_e -\frac{1}{3}[\Phi_e]$ 
into the potential \cite{e-yukawaon-PRD09}. 
In the present model, we also consider a more general 
form of $W_\Phi$, 
$$
W_\Phi= 
  \varepsilon_{SB} \lambda_A \left\{ \alpha_1 
[\hat{\Phi}_e \hat{\Phi}_e \hat{Y}_e] 
+\alpha_2 [\Phi_e \Phi_e \hat{Y}_e]
+\alpha_3 [\Phi_e \Phi_e Y_e] \right\}
$$
$$
+ \varepsilon_{SB}  \lambda_B \left\{ \beta_1 
[\hat{\Phi}_e \hat{\Phi}_e \hat{\Phi}_e] 
+\beta_2 [\Phi_e \Phi_e \hat{\Phi}_e]
+\beta_3 [\Phi_e \Phi_e \Phi_e] \right\},
\eqno(2.1)
$$
where $\hat{\Phi}_e$ and $\hat{Y}_e$ are ${\bf 5}$ 
components of O(3) which are defined by
$$
\hat{\Phi}_e \equiv \Phi_e -\frac{1}{3}[\Phi_e] , \ \ \ 
\hat{Y}_e \equiv Y_e -\frac{1}{3}[Y_e].
\eqno(2.2)
$$
(Hereafter, we will drop the index ``$e$", because 
we only deal with fields in the charged lepton sector 
in this paper.)
The expression (2.1) is the most general form of 
$\Phi \Phi Y$ and $\Phi \Phi \Phi$ terms, because we know 
the following identical equations:
$$
[\Phi \hat{\Phi} \hat{Y}]= \frac{1}{2} \left(
[\hat{\Phi} \hat{\Phi} \hat{Y}] + [{\Phi} {\Phi} \hat{Y}]
\right) , 
\eqno(2.3)
$$
$$
[\hat{\Phi} {\Phi} {Y}] = [\hat{\Phi} \hat{\Phi} {Y}]
+ [{\Phi} \hat{\Phi} \hat{Y}] -[\hat{\Phi} \hat{\Phi} \hat{Y}] ,
\eqno(2.4)
$$
and an equation with $Y \rightarrow \Phi$ in Eq.(2.3).
We consider that the parameters $\lambda_e$, 
$\lambda_A$, $\lambda_B$, $\mu_e$ and $\varepsilon_{SB}$
in the superpotential are ``analogue" parameters 
which we can adjust continuously, 
while we assume that $\alpha_i(\Lambda)$ and $\beta_i(\Lambda)$ 
should be integers. 
We consider that the relations for $K(\Lambda)$ and 
$\kappa(\Lambda)$ are independent of these analogue
parameters, and those should be described only by
the ``digital" parameters $\alpha_i$ and $\beta_i$.
Hereafter, we will simply denote $K(\Lambda)$,  
$\kappa(\Lambda)$, $\dots$, as $K$, $\kappa$, $\cdots$,
because we discuss only the relations among quantities
at $\mu=\Lambda$.

The terms given in Eq.(2.1) explicitly
break not only the U(1)$_X$ symmetry but also the
$R$ symmetry, because we assign the $R$ charges of 
$\Phi$, $Y$ and $\Theta$ to $0$, $0$ and $2$, 
respectively.
We assume that the value $\varepsilon_{SB} $ is negligibly 
small.
Therefore, the present model cannot break the supersymmetry
spontaneously \cite{R-sym}.
For the moment, we consider that the supersymmetry is
unbroken, at least, in the yukawaon sector.

The VEV of the ur-yukawaon $\Phi$ is obtained from
SUSY vacuum conditions as we show later. 
Then, we will obtain a cubic equation for  
eigenvalues $v_i$ of $\langle\Phi\rangle$:
$$
c_3 \langle\Phi\rangle^3 + c_2 \langle\Phi\rangle^2 
+c_1 \langle\Phi\rangle + c_0 {\bf 1} = 0 ,
\eqno(2.5)
$$
which necessarily and sufficiently determines the 
eigenvalues of  $\langle\Phi\rangle$.
The coefficients $c_a$ ($a=0,1,2,3$), in general, have 
the following relations:
$$
\frac{c_2}{c_3}= - [\langle\Phi\rangle] , \ \ \
\frac{c_1}{c_3} = \frac{1}{2} \left( [\langle\Phi\rangle]^2 
-[\langle\Phi\rangle\langle\Phi\rangle] \right) , \ \ \
\frac{c_0}{c_3} = -\det \langle\Phi\rangle .
\eqno(2.6)
$$
Therefore, we can express the quantities $K$ and $\kappa$ 
as follows:
$$
K = 1 -2 \frac{c_1}{c_3} \frac{1}{[\langle\Phi\rangle]^2} ,
\eqno(2.7)
$$
$$
\kappa = -\frac{c_0}{c_3} \frac{1}{[\langle\Phi\rangle]^3} ,
\eqno(2.8)
$$
by using the relations (2.6).

Prior to calculating SUSY vacuum conditions, it is useful
to the expression (2.1) as follow:
$$
W_\Phi= 
 \varepsilon_{SB}  \lambda_A \left\{ 
a_0 [\Phi \Phi Y] +a_1 [\Phi] [\Phi Y] 
+a_2 [\Phi]^2 [Y] +a_3 [\Phi \Phi] [Y] \right\}
$$
$$
+ \varepsilon_{SB}  \lambda_B \left\{  
b_0 [\Phi \Phi \Phi] + b_1 [\Phi \Phi] [\Phi]
+b_2 [\Phi]^3 \right\},
\eqno(2.9)
$$
where
$$
a_0=\alpha_1+\alpha_2+\alpha_3, \ \ 
a_1=-\frac{2}{3} \alpha_1, \ \ 
a_2=\frac{2}{9} \alpha_1, \ \ 
a_3=-\frac{1}{3} (\alpha_1+\alpha_2), 
\eqno(2.10)
$$ 
$$
b_0=\beta_1+\beta_2+\beta_3, \ \ 
b_1=-\frac{1}{3} (3\beta_1+\beta_2) , \ \ 
b_2=\frac{2}{9} \beta_1. 
\eqno(2.11)
$$ 
Then, from a SUSY vacuum condition 
$\partial W/\partial Y=0$, we obtain
$$
\Theta = -\varepsilon_{SB}  \frac{\lambda_A}{\mu}
\left\{ a_0 \Phi \Phi + a_1 [\Phi] \Phi 
+a_2 [\Phi]^2 {\bf 1} + a_3 [\Phi \Phi] {\bf 1}
\right\} .
\eqno(2.12)
$$
From a SUSY vacuum condition
$\partial W/\partial \Phi=0$, we also obtain
$$
\frac{\partial W}{\partial \Phi}=0 = \lambda 
(\Phi \Theta +\Theta \Phi) 
$$
$$
+ \varepsilon_{SB}  \lambda_A \left\{ a_0 (\Phi Y+Y\Phi)
+a_1 [\Phi] Y + a_1 [\Phi Y] {\bf 1} 
+ 2 a_2 [Y][\Phi]{\bf 1} + 2 a_3 [Y]\Phi \right\}
$$
$$
+ \varepsilon_{SB}  \lambda_B \left\{ 3 b_0 \Phi\Phi 
+ 2 b_1 [\Phi]\Phi +b_1 [\Phi\Phi] {\bf 1} 
+ 3 b_2[\Phi]^2 {\bf 1} \right\} .
\eqno(2.13)
$$
When the value of $\varepsilon_{SB} $ is small, but it is 
not zero, by substituting Eqs.(1.16) and (2.12) for Eq.(2.13), 
we obtain the cubic equation (2.5) with coefficients
$$
c_3= 4 a_0, \ \ c_2= (3 a_1 -\xi b_0)[\Phi] , 
\eqno(2.14)
$$
$$
c_1 = 2(a_2 [\Phi]^2+2 a_3 [\Phi\Phi] -\xi b_1 [\Phi]^2),
\eqno(2.15)
$$
$$
c_0 = a_1 [\Phi\Phi\Phi] +2 a_2 [\Phi\Phi][\Phi]
-\xi [\Phi](b_1[\Phi\Phi]+3b_2[\Phi]^2) ,
\eqno(2.16)
$$
where $\xi$ is defined by
$$
\xi = \frac{\lambda_B}{\lambda_A} \frac{\mu}{\lambda [\Phi]},
\eqno(2.17)
$$
and it is an ``analogue" parameter at present.
Note that the eigenvalues of $\langle \Phi\rangle$ are 
independent of the analogue parameters except for $\xi$
(however, we will show later that $\xi$ becomes  
a digital one).

By using a general formula for any $3\times 3$ matrix
$$
L \equiv \frac{[\Phi\Phi\Phi]}{[\Phi]^3} =
3 \kappa + \frac{3}{2} K - \frac{1}{2} ,
\eqno(2.18)
$$
together with Eqs.(2.7) and (2.8), 
we can obtain
$$
K = 1- 2 \frac{c_1}{c_3} \frac{1}{[\Phi]^2} 
= 1 -\frac{1}{a_0}(a_2 + 2 a_3 K -\xi b_1) ,
\eqno(2.19)
$$
i.e.
$$
K= \frac{ 1-\frac{a_2}{a_0} + \xi \frac{b_1}{a_0} }{
1+2 \frac{a_3}{a_0} } ,
\eqno(2.20)
$$
and
$$
\kappa = -\frac{c_0}{c_3} \frac{1}{[\Phi]^3} 
= -\frac{a_1}{4a_0}\left\{3 \kappa +\frac{3}{2} K -\frac{1}{2} 
+2 \frac{a_2}{a_1} K - 2 \xi\frac{b_1}{a_1}\left( K
+ 3\frac{b_2}{b_1} \right) \right\} ,
\eqno(2.21)
$$
i.e.
$$
\kappa = \frac{1}{1+\frac{3 a_1}{4 a_0} }
\frac{a_1}{8a_0} \left\{ 1-3 K -4\frac{a_2}{a_1}K
+2 \xi  \frac{b_1}{a_1} \left(K + 3\frac{b_2}{b_1}\right) 
\right\}.
\eqno(2.22)
$$

On the other hand, as seen in (2.6), in order to obtain 
a solution with $[\Phi]\neq 0$, from the relations (2.14),  
the parameter $\xi$ must satisfy 
$$
0=c_3 + \frac{c_2}{[\Phi]} =4 a_0 + 3 a_1 -3 \xi b_0  .
\eqno(2.23)
$$
Therefore, the parameter $\xi$ is not an adjustable 
parameters, and, now, it is expressed by 
the digital parameters $\alpha_i$ and $\beta_i$. 

When we define a deviation of $\varepsilon$ from 
the ideal value $K=2/3$ as
$$
K= \frac{2}{3}(1+\varepsilon) ,
\eqno(2.24)
$$
the quantities $\kappa$ and $K$ are expressed as 
follows: 
By using Eqs.(2.10) and (2.11) together with Eq.(2.23),
the quantity $\kappa$ is expressed as 
$$
\kappa =\frac{1}{54 \rho} ( 1-12 \rho \sigma + 2 \omega
\varepsilon ) ,
\eqno(2.25)
$$
where
$$
\rho = -\left(1+\frac{4 a_0}{3 a_1} \right) =
\frac{\alpha_1+ 2\alpha_2 +2\alpha_3}{\alpha_1} ,
\eqno(2.26)
$$
$$
\sigma = -\left( 1+\frac{9 b_2}{2 b_1}\right)
\frac{b_1}{b_0} = \frac{1}{3} 
\frac{\beta_2}{\beta_1+\beta_2+\beta_3} ,
\eqno(2.27)
$$
$$
\omega = 5+6 \rho \frac{b_1}{b_0} =
5 -2 \frac{\alpha_1+ 2\alpha_2 +2\alpha_3}{\alpha_1}
\frac{3 \beta_1+\beta_2}{\beta_1+\beta_2+\beta_3} .
\eqno(2.28)
$$
Also, from Eq.(2.20), the deviation $\varepsilon$ is
expressed as
$$
\varepsilon = \frac{3}{2} K -1 =
\frac{\Delta }{2 (\alpha_1+ \alpha_2 +3\alpha_3)
(\beta_1+\beta_2+\beta_3) } ,
\eqno(2.29)
$$
where
$$
\Delta =2 ( \alpha_1+ 2\alpha_2 +2\alpha_3)(\beta_3
-2 \beta_1) -( 3 \alpha_1+ 3\alpha_2 -\alpha_3)
(\beta_1+\beta_2+\beta_3) .
\eqno(2.30)
$$
Since $K(\Lambda)=\frac{2}{3}(1+\varepsilon(\Lambda))= 
K^{pole}(1+\varepsilon_K(\Lambda))$, 
i.e. $K^{pole}=\frac{2}{3}(1+\varepsilon(\Lambda) 
-\varepsilon_K(\Lambda))$, 
we will search for a model which gives
$\varepsilon(\Lambda) \simeq \varepsilon_K(\Lambda)$ 
numerically.
(Note that this does not mean 
$\varepsilon(\mu) \simeq \varepsilon_K(\mu)$ at every energy
scale $\mu$.)


\vspace{3mm}

\noindent{\large\bf 3 \ Toy model with a simpler structure}

In this section, we investigate a model with a simpler form
$$
W_\Phi= 
 + \varepsilon_{SB} \lambda_A \left\{ \alpha_1 
[\hat{\Phi} \hat{\Phi} \hat{Y}] 
+\alpha_3 [\Phi \Phi Y] \right\}
+ \varepsilon_{SB} \lambda_B \left\{ \beta_1 
[\hat{\Phi} \hat{\Phi} \hat{\Phi}] 
+\beta_3 [\Phi \Phi \Phi] \right\},
\eqno(3.1)
$$
i.e. with a case $\alpha_2=\beta_2=0$ in Eq.(2.1).
Here, we assume that $\alpha_1$, $\alpha_3$, 
$\beta_1$ and $\beta_3$ are positive integers.
(Note that this ansatz is only valid at $\mu=\Lambda$.)
In the present model, we essentially have two 
parameters $\alpha_3/\alpha_1$ and $\beta_3/\beta_1$
for the two observables $K(\Lambda)$ and $\kappa(\Lambda)$,
so that we cannot predict $K$ and $\kappa$ definitely.
For large values $\alpha_i$ and $\beta_i$, the 
parameters $\alpha_3/\alpha_1$ and $\beta_3/\beta_1$
approximately behave as if they are continuous 
parameters.
In order to exclude such the cases from consideration, 
we assume that $\alpha_i$ and $\beta_i$
are not so large, e.g. they should be smaller than
about 50.
(When the parameters $\alpha_1$ and $\alpha_3$ are 
given by $\alpha_1= k n_1$ and $\alpha_3=k n_3$ 
($k$, $n_1$ and $n_3$ are positive integers), we 
redefine $\alpha_1$ and $\alpha_3$ as $\alpha_1=n_1$
and $\alpha_3=n_3$.
We also do such a re-definition for the parameters
$\beta_1$ and $\beta_3$.)

In the present model (3.1), since $\beta_2=0$, the 
factor $\sigma$ defined in (2.27) becomes $\sigma=0$,
so that the quantity $\kappa$ is expressed simply as
$$
\kappa = \frac{1}{54 \rho} (1+2\omega \varepsilon) .
\eqno(3.2)
$$
Since $\varepsilon \sim 10^{-3}$, the value $\kappa$
is almost determined by the parameter $\alpha_3/\alpha_1$:
$$
\rho =1 + 2 \frac{\alpha_3}{\alpha_1} .
\eqno(3.3)
$$
Since we search a case $\kappa \neq 0$, we consider
$\alpha_1 \neq 0$.
Since $\partial \kappa/\partial \alpha_1 >0$, 
we can regard the value of $\alpha_1$ as 
$\alpha_1=1$, which gives a next smaller value of
$\kappa$ following to $\kappa=0$ at $\alpha_1=0$.
Then, by neglecting the factor $2\omega\varepsilon
\lesssim 10^{-2}$, the expression (3.2) prdicts
$$
\kappa \simeq \frac{1}{54(1+2\alpha_3)} .
\eqno(3.4)
$$
On the other hand, the deviation $\varepsilon$ 
defined by (2.24) is given by
$$
\varepsilon = \frac{\Delta}{2(3\alpha_3+1)(\beta_3+\beta_1)},
\eqno(3.5)
$$
where
$$
\Delta =(3 \alpha_3+5)\beta_3 - (9\alpha_3 +1) \beta_1.
\eqno(3.6)
$$
The smallest value of $\varepsilon$ for each $\alpha_3$ can 
appear when $\Delta=1$.
(The equation $\Delta=1$ is known as a Diophantine equation.)
Our interest is whether there is a simpler set of
$(\alpha_3, \beta_1, \beta_3)$ which can favorably give
the observed values $K(\Lambda)$ and $\kappa(\Lambda)$.

Obviously, from Eq.(3.6), when $\beta_1 =\beta_1^0 \equiv
3 \alpha_3+5$ and  $\beta_3 =\beta_3^0 \equiv
9 \alpha_3+1$, the case $\beta\equiv (\beta_1,\beta_3) 
=\beta_0 \equiv (\beta_1^0, \beta_3^0)$ gives 
$\varepsilon = 0$.
[Exactly speaking, the lowest number of 
$\beta_0 \equiv (\beta_1^0, \beta_3^0)$  which gives 
$\varepsilon=0$ is given by 
$(\frac{1}{2}(3\alpha_3+5), \frac{1}{2}(9\alpha_3+1))$
for an odd number of $\alpha_3$, and 
$\beta_0=(3k+2, 9k+4)$ for $\alpha_3=3 \mod 7$ 
(i.e. for $\alpha_3=7k+3$).]
Since we want a case with $\varepsilon \neq 0$, we 
consider $\beta \neq \beta_0$.
For $\beta >\beta_0$, i.e. for large values of
$\beta_1$ and $\beta_3$, we can obtain any small 
value of $\varepsilon$.
In order to exclude such the cases from consideration, 
we assume that the values $\beta_1$ and $\beta_3$
are not so large, and we confine our value 
$\beta=(\beta_1,\beta_3)$ as
$$
(1,1) \leq (\beta_1,\beta_3) < (3\alpha_3+5, 9\alpha_3+1) .
\eqno(3.7)
$$
At $\beta_1=\beta_1^0$, $\varepsilon$ takes $\varepsilon<0$
for $1 \leq \beta_3 \leq \beta_3^0-1$ and $\varepsilon = 0$
for $\beta_3 =\beta_3^0$.
At the next value $\beta_1=\beta_1^0-1$, $\varepsilon$ takes 
$\varepsilon<0$ for $1 \leq \beta_3 \leq \beta_3^-$ and 
$\varepsilon >0$ for $\beta_3^- +1 \leq \beta_3 \leq \beta_3^0$. 
(For the explicit value of $\beta_3^-$, see Table 2. 
In Table 2, a value of $\beta_3$ in each $\varepsilon_{(n)}$ 
denotes $\beta_3=\beta_3^- +1$.)
In general, for a given value $\beta_1 = \beta_1^0 -n$,
the value of $\varepsilon$ takes the smallest value 
$\varepsilon_{(n)}$ at $\beta_3 =\beta_3^-+1$.
We list the values $\varepsilon_{(1)}$ and $\varepsilon_{(2)}$
for typical cases of $\alpha=(\alpha_1,\alpha_3)$ and 
$\beta=(\beta_1,\beta_3)$ in Table 2.
Note that the array $(\varepsilon_{(1)}, \varepsilon_{(2)},
\dots )$ are not always in order of magnitude.
For example, the smallest value in $\alpha_3=4$ is 
$\varepsilon_{(6)}=1/910$.


\begin{table}
\caption{
Smallest value of $\varepsilon_{(n)}$ for $\beta_1=\beta_1^0 -n$.
$(\beta_1^0, \beta_3^0)$ denotes a value which gives 
$\varepsilon=0$ for a given value of $\alpha_3$.}
\vspace{1mm}

\begin{tabular}{|c|c|c|c|c|c|c|c|c|c|c|} \hline
$\alpha_3$ & $1$ & $2$ & $3$ & $4$ & $5$ & $6$ & $7$ &
$8$ & $9$ & $10$ \\ \hline
$(\beta_1^0,\beta_3^0)$ & $(4,5)$ & $(11,19)$ & 
$(1,2)$ & $(17,37)$ & $(10,23)$ & $(23,55)$ & 
$(13,32)$ & $(29,73)$ & $(16,41)$ & $(5,13)$ \\ \hline
$\varepsilon_{(1)}$ & $1/28$ & $1/49$ & --- & $1/442$ &
$1/160$ & $3/950$ & $1/154$ & $1/330$ & $1/168$ & $7/310$ \\
$(\beta_1,\beta_3)$ & $(3,4)$ & $(10,18)$ & --- & 
$(16,35)$ & $(9,21)$ & $(22,53)$ & $(12,30)$ & $(28,71)$ &
$(15,39)$ & $(4,11)$ \\ \hline
$\varepsilon_{(2)}$ & $1/10$ & $1/70$ & --- & $1/208$ &
$1/72$ & $1/152$ & $2/143$ & $1/4750$ & $1/700$ & $7/682$ \\
$(\beta_1,\beta_3)$ & $(2,3)$ & $(9,16)$ & --- & 
$(15,33)$ & $(8,19)$ & $(21,51)$ & $(11,28)$ & $(27,68)$ &
$(14,36)$ & $(3,8)$ \\ \hline

\end{tabular}
\end{table}

In Eq.(3.4), we have assumed $\alpha_1=1$ as the next
value to $\alpha_1=0$ which gives $\kappa=0$, because
we consider $\kappa \neq 0$ in the realistic world.
Similarly, we assume $\beta_1=\beta_1^0 -1$ as the next
value to $\beta_1=\beta_1^0$ which gives $\varepsilon=0$,
because we consider $\varepsilon \neq 0$ in the realistic
world.
(Since $\partial \varepsilon/\partial \beta_1 <0$ against
$\partial \kappa/\partial \alpha_1 >0$, we take the next
value as $\beta_1=\beta_1^0 -1$ against $\alpha_1= 0 +1$.)
We assume that the physical value of $\varepsilon$ is 
given by the smallest value $\varepsilon_{(1)}$ at 
$\beta=(\beta_1^0-1,\beta_3^- +1)$.  
As seen in Table 2, of the smallest values of $\varepsilon_{(1)}$
for various values of $\alpha_3$, the smallest one is 
$$
\varepsilon =\frac{1}{442} = 2.26244 \times 10^{-3} ,
\eqno(3.8)
$$
at $\alpha=(1,4)$ and 
$\beta=(16,35)$.
The value (3.8) is in good agreement with the value of 
$\varepsilon_K$ given in (1.12).
(Exactly speaking, the value (3.8) is in excellent agreement
with the value of $\varepsilon_K$ at $\mu \sim 10^{11}$ GeV
as seen in Table 1, although we consider $\Lambda \sim 10^{15}$
GeV from a naive estimate $m_\nu \sim \langle H_u^0\rangle^2/\Lambda$.)
On the other hand, the choice $\alpha=(1,4)$ predicts
$$
\kappa = \frac{1}{486} (1+ 2 \omega \varepsilon) ,
\eqno(3.9)
$$
where $\omega = -{609}/{51} = - 11.94$.
By inputting the value (3.8) into Eq.(3.9), we obtain 
$$
\kappa = 2.0756\times 10^{-3} \times (1-0.0540)
= 1.9464 \times 10^{-3}.
\eqno(3.10)
$$

It is interesting that the value of $\kappa$ in the limit
of $\varepsilon=0$ in Eq.(3.9) 
$$
\kappa|_{\varepsilon=0} = \frac{1}{486} = 2.058\times 10^{-3} ,
\eqno(3.11)
$$
is in good agreement with the observed value 
$\kappa^{pole}=(2.0633\pm 0.0001)$ correspondingly to
the fact that $K$ in the limit of $\varepsilon=0$ gives
$K=2/3=K^{pole}$.
(The value $\kappa=1/486$ has first been speculated from
another scenario \cite{Koide-U3-PLB08}.)
However, $K$ and $\kappa$ in the present section denote
$K(\Lambda)$ and $\kappa(\Lambda)$, respectively, so that 
there is no theoretical ground for the conjecture that
$K$ and $\kappa$ in the limit $\varepsilon=0$ give 
$K^{pole}$ and $\kappa^{pole}$.


\vspace{3mm}

\noindent{\large\bf 4 \ Concluding remarks}

First, we would like to re-emphasize that  the quantities $K(\mu)$
and $\kappa(\mu)$, which are defined by Eqs.(1.2) and (1.3), 
respectively, are useful for investigating the charged lepton mass
spectrum, because they are almost independent of the energy scale
$\mu$ of the running masses.

In the present paper, against a conventional view, we have taken 
a standpoint that the observed relation $K^{pole}=2/3$ 
is merely an approximate relation, and 
the ratio $K$ is given by 
$K(\Lambda)=\frac{2}{3}(1+\varepsilon(\Lambda))$
with $\varepsilon(\Lambda) \sim 10^{-3}$ 
(not $\varepsilon(\Lambda) = 0$) from the beginning.  
In order to understand  $\varepsilon(\Lambda)\sim 10^{-3}$, 
on the basis of the so-called supersymmetric yukawaon model, 
we have investigated possible values of $K(\Lambda)$ and 
$\kappa(\Lambda)$.
By taking a simpler form of the superpotential of the 
yukawaons (3.1) at $\mu=\Lambda$,  and by assuming that the 
coefficient of each term in (3.1) is given by a lower positive 
integer at $\mu=\Lambda$, we have tried to speculate values of 
$K(\Lambda)$ and $\kappa(\Lambda)$.
However, the present model (3.1) involves 
two parameters $\alpha_3/\alpha_1$ and $\beta_3/\beta_1$,
so that the model cannot predict values of $K(\Lambda)$ and 
$\kappa(\Lambda)$ definitely.
The model can predict only many candidates of the values
$K(\Lambda)$ and $\kappa(\Lambda)$ (but as discrete values).
Of the many candidates of the predicted values, a case
of $\alpha=(1,4)$ and $\beta=(16,35)$ (i.e. 
$\alpha_3/\alpha_1=4$ and $\beta_3/\beta_1=2+3/16$) can 
give interesting results of $K(\Lambda)$ and $\kappa(\Lambda)$:
$\varepsilon = 1/442=2.262\times 10^{-3}$  and 
$\kappa =(1/486)\times (1+2\omega\varepsilon)
=2.058\times 10^{-3}\times (1-0.054)=1.95 \times 10^{-3}$,
which are in fairly good agreement with our goal  
$\varepsilon_K(\Lambda) \simeq 2.4 \times 10^{-3}$ and 
$\kappa(\Lambda) \simeq 2.02\times 10^{-3}$, respectively.
Although, on trial in Sec.3, we have assumed that the 
nature chooses the smallest $\varepsilon$ in the next 
$\beta_1$ to $\beta_1=\beta_1^0$ (i.e. $\beta_1=\beta_1^0-1$),
and picked up the case of $\alpha=(1,4)$ and $\beta=(16,35)$,
there is no theoretical ground for such a selection rule.

The present model is based on an effective theory which is
valid at $\mu \leq \Lambda$. 
We have assumed that there is such the energy scale $\Lambda$ 
at which the superpotential takes a simple form. 
A naive estimate of $\Lambda$ leads to 
$m_\nu \sim \langle H_u^0\rangle^2/\Lambda$, so that 
we consider $\Lambda\sim 10^{15}$ GeV.
However, by taking VEV values of yukawaons and coefficients
(coupling constants) suitably, we may consider 
$\Lambda\sim 10^{14-16}$ GeV. 
As we have emphasized in Sec.1, the effective coupling 
constants $Y_f^{eff}= (y_f/\Lambda)\langle Y_f\rangle$
evolve as in the standard model below the scale 
$\Lambda$. 
Although the renormalization group equations (RGEs) affect 
to the parts ${\bf 1}$ and ${\bf 5}$ of yukawaons with
$({\bf 3}\times{\bf 3})_S$ of O(3) separately, 
the effects are negligible in the present model.
(For example, see Sec.6 in Ref.\cite{Yamashita09}.)  

Exactly speaking, a threshold effect at $\mu\sim \Lambda$ is 
also important in the present model.  The singlet and  5-plet 
parts of the effective Yukawa interactions will separately 
receive such corrections in the present effective theory.  
However, in order to estimate the effect exactly, we must 
know a full theory at $\mu > \Lambda$.  Since we do not know 
such the full theory at present, in this paper, we are obliged 
to put an ansatz that the threshold effect is negligibly small.  
Our goal is to simultaneously understand the both values of 
$K(\Lambda)$ and $\kappa(\Lambda)$ at $\mu=\Lambda \sim 10^{14-16}$ 
GeV which are given in Table 1, even at the cost of such the 
accuracy in the observed fact $K^{pole} \simeq 2/3$.  
  
The predicted values (3.8) and (3.10) are only valid within 
an approximation where we neglect such the threshold effect. 
In fact, the predicted value (3.8) at $\mu=\Lambda$ does not 
lead to $K^{pole}=2/3$ in the exact meaning.  

The superpotential terms $W_\Phi$ given in Eq.(3.1) 
spontaneously break the U(1)$_X$ symmetry, so that 
pseudo Goldstone bosons appear.
An earlier estimate of such light yukawaons based on
a prototype model of the present $W_\Phi$ is found in
Ref.\cite{y-mass-0902}.
The conclusions in Ref.\cite{y-mass-0902} that we have
three massless scalars and three pseudo Goldstone bosons 
with masses of the order of $\varepsilon_{SB} \Lambda$
will be unchanged in the present model.
For more details of physical meaning of the light bosons,
we will discuss elsewhere.

In this paper, we have considered that the observed fact
$K^{pole}=2/3$ is accidental, and the model should give 
$K(\Lambda)=\frac{2}{3}(1+\varepsilon(\Lambda))$ with 
$\varepsilon(\Lambda) \sim 10^{-3}$ from the beginning.
However, the present model have failed to give such excellent 
coincidences about $K(\Lambda)$ and $\kappa(\Lambda)$ as the 
well-known relation $K^{pole}=2/3$.
In contrast to the present idea, recently, Sumino 
\cite{Sumino09} has proposed a model which gives
$K^{pole}=2/3$ exactly. 
This idea is based on a point of view that the relation (1.1) 
for the pole masses is exactly valid, so that his standpoint
is in opposite direction against the present approach.
If the charged leptons are truly fundamental entities, the 
laws of the nature for the charged leptons should be 
beautifully simple, so that it seems to be likely that 
$K^{pole}=2/3$ is exact as pointed out by Sumino. 
Even if Sumino's idea is true, the present approach based 
on the supersymmetric yukawaon model will still be promising
for unified understanding of $K(\Lambda)$ and 
$\kappa(\Lambda)$, 
because the cubic equation (2.5) can completely determine 
the eigenvalues of $\langle\Phi\rangle$ (i.e. $K(\Lambda)$ 
and $\kappa(\Lambda)$ simultaneously).  
We hope that a simple requirement for $W_\Phi$
leads to simultaneous understanding of $K(\Lambda)$ 
and $\kappa(\Lambda)$ along a line suggested by a 
yukawaon model.

\vspace{6mm}

\centerline{\large\bf Acknowledgments}

The author would like to thank Y.~Hyakutake,
M.~Tanaka, N.~Uekusa for helpful discussions.
Especially, he thanks T.~Yamashita, T.~Onogi, Y.~Sumino 
and K.~Oda for valuable and helpful comments on the renormalization 
group equation effects and the threshold effects.
He also thanks Midori Kobayashi for valuable suggestions 
concerning a Diophantine equation. 
This work is supported by the Grant-in-Aid for
Scientific Research (C), JSPS, (No.21540266).


\end{document}